# A Gate-tunable Polarized Phase of Two-Dimensional Electrons at the LaAlO$_3$/SrTiO$_3$ Interface

Arjun Joshua, J. Ruhman, S. Pecker, E. Altman and S. Ilani

*Department of Condensed Matter Physics, Weizmann Institute of Science, Rehovot 76100, Israel.*

**Controlling the coupling between localized spins and itinerant electrons can lead to exotic magnetic states. A novel system featuring local magnetic moments and extended two-dimensional electrons is the interface between LaAlO$_3$ and SrTiO$_3$. The magnetism of the interface, however, was observed to be insensitive to the presence of these electrons and is believed to arise solely from extrinsic sources like oxygen vacancies and strain. Here we show the existence of unconventional electronic phases in the LaAlO$_3$/SrTiO$_3$ system pointing to an underlying tunable coupling between itinerant electrons and localized moments. Using anisotropic magnetoresistance and anomalous Hall effect measurements in a unique in-plane configuration, we identify two distinct phases in the space of carrier density and magnetic field. At high densities and fields the electronic system is strongly polarized and shows a response which is highly anisotropic along the crystalline directions. Surprisingly, below a density-dependent critical field the polarization and anisotropy vanish whereas the resistivity sharply rises. The unprecedented vanishing of the easy axes below a critical field is in sharp contrast with other coupled magnetic systems and indicates strong coupling with the moments that depends on the symmetry of the itinerant electrons. The observed interplay between the two phases indicates the nature of magnetism at the LaAlO$_3$/SrTiO$_3$ interface as both having an intrinsic origin and being tunable.**

The electronic system at the LaAlO$_3$/SrTiO$_3$ (LAO/STO) interface (1) has shown an intriguing combination of superconductivity (2, 3), spin-orbit coupling (4, 5), and most recently, magnetism (6–13). An especially fascinating feature of this system is the existence of localized magnetic moments (14, 15) in proximity with itinerant d-electrons (16–21) resulting in interesting coexistence phenomena (7–10). An unresolved issue central to a microscopic understanding of these properties is whether the electrons and moments interact with each other. It was shown that the itinerant electrons can be gate-tuned through a Lifshitz transition (22), where they change from populating light $d_{XY}$ bands with a circular Fermi surface to occupying also heavy $d_{XZ}/d_{YZ}$ bands with highly-elongated elliptical Fermi surfaces oriented along crystalline axes. The latter bands can have preferred axes for anisotropy along crystalline directions (21). Preferred crystalline directionality may also arise due to the localized magnetic moments, since they too originate from d-orbitals localized on individual Ti atoms. Therefore, signatures of if and how the moments couple to the electrons will be embedded in the spatial character of the ground states of the LAO/STO system.

Measurements of anisotropic magnetoresistance (23) (AMR) in a rotating in-plane magnetic field are a powerful tool to determine these symmetries. Previous AMR measurements in this system have addressed the effects of surface terraces (24), possible magnetic ordering (25), and prominent Rashba spin-orbit interactions (26). Magnetic ordering in STO-based systems is also inferred from the anomalous Hall effect (AHE) in a perpendicular field (27). The interpretation of both AMR and AHE measurements at the LAO/STO interface, however, is complicated by a competing effect. On one hand, AMR measurements can be overwhelmed by orbital effects due to the slightest perpendicular field (25). Moreover, the multiband nature of conduction at the LAO/STO interface induces a nonlinear Hall effect thus mimicking the AHE even without any magnetization present (22, 28). On the other hand, direct scanning SQUID (29) and torque magnetometry (10) measurements show that the magnetization lies in-plane suggesting one probe for signatures of the interaction between the moments and the electrons in this specific geometry. In this work we use AMR with a high degree of alignment of the field to lie purely in the interfacial plane, in conjunction with measurements of AHE in the unconventional planar configuration, to probe the symmetries and polarization in this system. In the space of magnetic field and electron density we observe two distinct phases: The first is characterized by a weak non-crystalline AMR (where the AMR induced by the field does not depend on its direction with respect to the crystal axes), a normal Hall behavior, and a large longitudinal resistivity. The second region shows strong crystalline AMR (where the AMR depends on the orientation of the field with respect to the crystal axes), large AHE indicative of strong polarization, and a huge drop in longitudinal resistivity (4, 25). The transition between these regions occurs at a density-dependent critical field that diverges at the Lifshitz transition (where the shape of the Fermi surface changes from circular to elliptical as the chemical potential crosses into the $d_{XZ}/d_{YZ}$ bands), demonstrating the crucial role played by itinerant electrons in the observed phases. This unusual behavior cannot be explained by considering only the intrinsic energy bands or scattering by magnetic moments, but is shown to naturally follow from a model wherein both these components are correlated via strong coupling between them that changes sign depending on whether the electrons are of $d_{XY}$ or $d_{XZ}/d_{YZ}$ character.

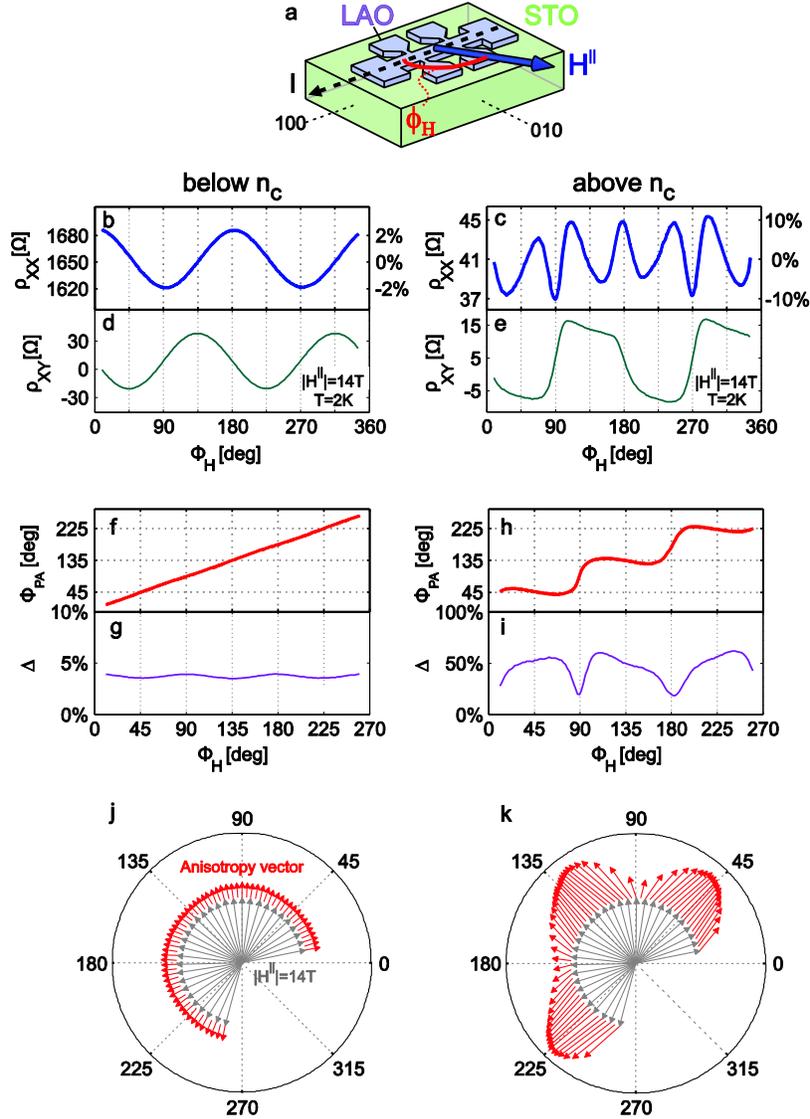

**Figure 1: Anisotropic Magneto Resistance (AMR) measurements below and above the Lifshitz critical density, $n_c$** a) A Hall bar along the (100) crystallographic direction in LAO/STO used for measuring the transport with in-plane magnetic field, $H^{\|}$, oriented at various angles, $\phi_H$, with respect to the current direction. b) and d) Measured longitudinal resistivity, $\rho_{XX}$, and transverse resistivity, $\rho_{XY}$, for $|H^{\|}| = 14T$ as a function of $\phi_H$, at a gate voltage of $V_g = 20V$, corresponding to a total carrier density $n = 1.58 \cdot 10^{13} \, cm^{-2}$, just below the Lifshitz transition density in this sample (22), $n = 1.62 \cdot 10^{13} \, cm^{-2}$. c) and e) Similar measurements for $V_g = 280V$, corresponding to a total density, $n = 2.3 \cdot 10^{13} \, cm^{-2}$, which is above $n_c$. The relative change in $\rho_{XX}$ is indicated on the right y-axes. f) and g) Direction of the principal axes of the anisotropy with respect to the current, $\phi_{PA}$, and its magnitude, $\Delta$, extracted by diagonalizing the resistivity tensor from the data below $n_c$ shown in panel b, d (see text). h) and i) Similar results for the data above $n_c$ (panels c, e). A small offset of 3.8 Ω was removed from $\rho_{XY}$ and $\rho_{YX}$ to make them symmetric around zero. Similar analysis without the offset removed also gives pinning of the anisotropy along diagonal directions (as in panel h) but further breaks the symmetry between the (110) and (1$\bar{1}$0) directions. j) Anisotropy vector (red arrows) below $n_c$ determined by $\phi_{PA}$ and $\Delta$, for various in-plane angles $\phi_H$ of $|H^{\parallel}| = 14T$ (grey arrows). Note that for clarity the magnitude of the anisotropy vector has been scaled up by a factor of 4 compared to k) showing the corresponding results above $n_c$.

We observed similar behavior in two independent samples with 6uc and 10uc of LAO. Data from the first sample is presented in detail below (see Supplementary A6 for sample growth and processing details). The longitudinal and transverse resistivities ($\rho_{XX}$ and $\rho_{XY}$) were measured using Hall bars while rotating the sample in a magnetic field applied in the plane of the interface (Fig. 1a) at temperatures of T=2K. Special care has been taken to minimize the wobble in our rotation apparatus, since the small wobble of standard cryogenic rotators $(\sim 1°)$ produces a spurious perpendicular field component that oscillates in-sync with the angle of the field in the plane. $\rho_{XX}$ in LAO/STO being extremely sensitive to even small perpendicular fields (25), such wobble induces spurious $\rho_{XX}$ modulations that overwhelm the intrinsic in-plane field modulations that we wish to measure. To eliminate this artifact we constructed an especially low-wobble rotator apparatus ($<0.006°$) based on an Attocube piezo rotator (ANR200), and have taken special care to mount the sample on it with parallelism $< 0.1°$. The results reported in this paper are therefore free of the spurious artifacts due to perpendicular fields.

When we measure the AMR at large magnetic fields we observe a fundamental difference below and above the Lifshitz point. Figure 1b shows the longitudinal resistivity, $\rho_{XX}$, measured at a large magnetic field ($H = 14T$) as a function of the angle of the field in the plane, $\phi_H$, at a carrier density below the sample's Lifshitz critical density, $n_c = 1.62 \cdot 10^{13} cm^{-2}$ (see caption). At this density, $\rho_{XX}$ has a small modulation as a function of $\phi_H$ (~4%) that accurately follows a simple $\cos(2\phi_H)$ dependence (see also ref. (26)). The situation is quite different above $n_C$ (Fig. 1c), where the modulation is much larger (~20%), and has a complex angular dependence (25), which peaks and dips along special angular directions ($\phi_H = 90°, 180°, 270°$), besides subsidiary features at intermediate angles.

We also measure a surprisingly large off-diagonal resistivity, $\rho_{XY}$. Below $n_C$, $\rho_{XY}$ shows a simple dependence on $\phi_H$ (Fig. 1d), similar to $\rho_{XX}$, but shifted by 45 degrees ($\sim \sin(2\phi_H)$) with almost identical peak-to-peak modulation (~60Ω). Above $n_C$ (Fig. 1e), $\rho_{XY}$ modulations become square-wave-like with values comparable even to the average value of $\rho_{XX}$, suggesting that these two quantities should be considered on equal footing. It is important to note that $\rho_{XY}$ shown here is not related to a Hall effect: First, it is measured with precisely in-plane field and second, whereas the Hall effect $\rho_{XY}$ must be anti-symmetric in magnetic field and under exchange of the spatial coordinates $(x \leftrightarrow y)$, the measured $\rho_{XY}$ is symmetric in both.

The observed symmetric $\rho_{XY}$ is in fact a direct signature of the anisotropy in this system. A two-dimensional anisotropic system is fully characterized by a 2x2 resistivity tensor with principal axes along two orthogonal directions in the plane, along which the resistivity assumes its highest $(\rho_h)$ and lowest $(\rho_l)$ values (23). For a general angle between the direction of the current and that of the principal axis, $\phi = \phi_I - \phi_{PA}$, the full resistivity tensor reads:

$$\begin{bmatrix} \rho_{XX} & \rho_{XY} \\ \rho_{YX} & \rho_{YY} \end{bmatrix} = \rho_{XX}^{av} \begin{bmatrix} 1 + \Delta/2 \cdot \cos(2\phi) & \Delta/2 \cdot \sin(2\phi) \\ \Delta/2 \cdot \sin(2\phi) & 1 - \Delta/2 \cdot \cos(2\phi) \end{bmatrix} \quad \text{(Eq. 1)}$$

where $\rho_{XX}^{av} = (\rho_h + \rho_l)/2$ is the angle-averaged longitudinal resistivity, and $\Delta = (\rho_h - \rho_l)/\rho_{XX}^{av}$ is the relative magnitude of the anisotropy. Clearly, $\rho_{XY}$ is non-zero only if there is anisotropy present, i.e. $\rho_h \neq \rho_l$.

Below $n_C$ the data (Fig. 1b, d) corresponds to an anisotropy whose principal axis is determined solely by the direction of $H$ (i.e. $\phi_{PA} = \phi_H$), hence we term this a non-crystalline anisotropy. In this case, Eq. 1 reduces to simple cosine and sine dependencies:

$$\rho_{XX} = \rho_{XX}^{av}(1 + \Delta/2 \cdot \cos(2\phi_H)), \quad \rho_{XY} = -\rho_{XX}^{av}\Delta/2 \cdot \sin(2\phi_H), \quad \text{(Eq. 2)}$$

accurately capturing the 45° phase shift between $\rho_{XX}$ and $\rho_{XY}$, and their identical peak-to-peak amplitudes, as seen in our data below $n_C$. Any angular dependence that deviates from these two simple relations (Eq. 2) necessarily implies the existence of an additional direction that, together with the direction of $H$, determines the principal axis. The most natural direction is given by the underlying crystal. The $\phi_H$-dependence above $n_C$ therefore corresponds to crystalline anisotropy, namely, one in which the electronic system is affected by the existence of preferred crystalline directions.

In our experiments the direction of the current is fixed along the crystal axis, but we can still determine the directionality and magnitude of the anisotropy for each $H$, by knowing the corresponding four components of the resistivity tensor. We measure $\rho_{XX}, \rho_{XY}$ and $\rho_{YX}$ for every $\phi_H$ and derive $\rho_{YY}$ by assuming that the system has square symmetry in the plane and thus is invariant under reflection about, say $\phi_H = 135°$, yielding $\rho_{YY}(\phi_H) = \rho_{XX}(270 - \phi_H)$. By determining the eigenvectors and eigenvalues of the full resistivity tensor, after removing a small constant offset in $\rho_{XY}$ and $\rho_{YX}$ (Fig. 1 caption), we extract for every $\phi_H$ the direction of the principal axis of the anisotropy, $\phi_{PA}$, and its magnitude, $\Delta$. Below $n_C$, we find that the anisotropy is along $H$ ($\phi_{PA} \approx \phi_H$, Fig. 1f), and its magnitude is almost independent of $\phi_H$ (Fig. 1g), consistent with non-crystalline symmetry. Above $n_C$, $\phi_{PA}$ does not simply follow $\phi_H$, but rather gets pinned along diagonal crystalline directions (Fig. 1h). The overall magnitude of the anisotropy (~50%) is also ten-fold larger and depends on $\phi_H$, being enhanced when the field is away from the crystalline axes (Fig. 1i). This striking change in the nature of the anisotropy across the Lifshitz point is summarized in Figs. 1j,k.

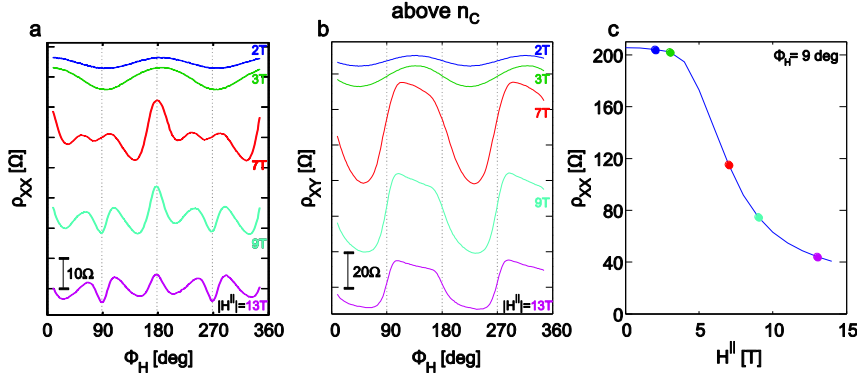

**Figure 2: Observation of a critical field in the AMR.** a) Longitudinal resistivity, $\rho_{XX}$, measured as a function of the angle of the field in the plane, $\phi_H$, at various field strengths (indicated). The curves were shifted along the y-axis for clarity (resistance scale is indicated on the bottom left). b) Corresponding $\rho_{XY}$ measurements. c) $\rho_{XX}$ as a function of the field strength for $\phi_H = 9°$. The colored dots mark the fields corresponding to the traces in panels a, b.

The change from non-crystalline to crystalline symmetry might be assigned to a change between $d_{XY}$ band occupation with an isotropic Fermi surface, to the population of $d_{XZ}/d_{YZ}$ orbitals with elliptical Fermi surfaces oriented along crystalline axes. On the other hand, the large square-wave-like angular dependence of $\rho_{XY}$ strongly resembles the anisotropy observed in semiconductors doped with magnetic impurities (30, 31). However there are fundamental differences between the LAO/STO system and magnetic semiconductors. These materials are intentionally doped with magnetic impurities whereas the local magnetic moments in LAO/STO are uncontrolled and their nature is still poorly understood. Compared to magnetic semiconductors (32), the itinerant d-electrons in the LAO/STO system have a much more anisotropic bandstructure than the itinerant holes in magnetic semiconductors which are derived from p-bands and the d-electrons can have an order of magnitude larger effective mass (33) than the holes, leading to enhanced correlation effects in the LAO/STO system. Furthermore, spin-orbit splitting in the bandstructure of LAO/STO is an order of magnitude smaller than that of the magnetic semiconductors. To better understand the possible interplay of magnetic moments and conduction electrons in the LAO/STO system we measured the field dependence of its AMR. Surprisingly, for densities well above $n_C$, where the $d_{xz}/d_{yz}$ bands are expected to be populated, the AMR at a *small* magnetic field is perfectly sinusoidal, namely, non-crystalline. Plotting the $\phi_H$-dependence of $\rho_{XX}$ (Fig. 2a) and $\rho_{XY}$ (Fig. 2b) for different magnetic fields we see a clear transition from non-crystalline to crystalline AMR, occurring at a critical field ($H_C^{\parallel} \approx 3T$, for the carrier density in Fig. 2). Furthermore, this change in AMR is concomitant with a huge fall (25) in $\rho_{XX}$ also commencing at $H_C^{\parallel}$ (Fig. 2c). The existence of a critical field cannot be explained by a single-particle band interpretation. It is also completely *opposite* to the trend seen in magnetic semiconductors where the AMR switches from crystalline to non-crystalline with increasing field (30, 31). Finally, in contrast to magnetic semiconductors where hysteresis is observed in $\rho_{XY}$ vs. $\phi_H$ due to switching of the easy axis (30, 31), we do not observe any such hysteresis.

Figure 3a maps out the magnitude of anisotropy in the space of electron density and in-plane magnetic field using the peak-to-peak modulation of $\rho_{XY}$ (see caption for details). Two distinct regions are clearly visible in the phase diagram: one with a small anisotropy ($\leq 4\%$, blue) and another with a large one (~50%, red). Within these regions, the magnitude of the anisotropy varies very little but at their boundary (dashed black line) it changes sharply. Interestingly, the $\phi_H$-averaged value of $\rho_{XX}$ changes throughout this phase diagram in perfect synchrony with the AMR (Fig. 3b): $\rho_{XX}$ is large in the region of small anisotropy and it drops to an asymptotic value about six-fold smaller in the region of large anisotropy.

The most striking feature in the phase diagram is that the critical field $H_C^{\parallel}$ continuously rises with decreasing density (dashed black line) and appears to diverge at the Lifshitz density (Figs. 3c and its inset). Indeed, below this critical density we do not observe crystalline AMR at all. Curiously, both the trend and the magnitude of $H_C^{\parallel}$ are

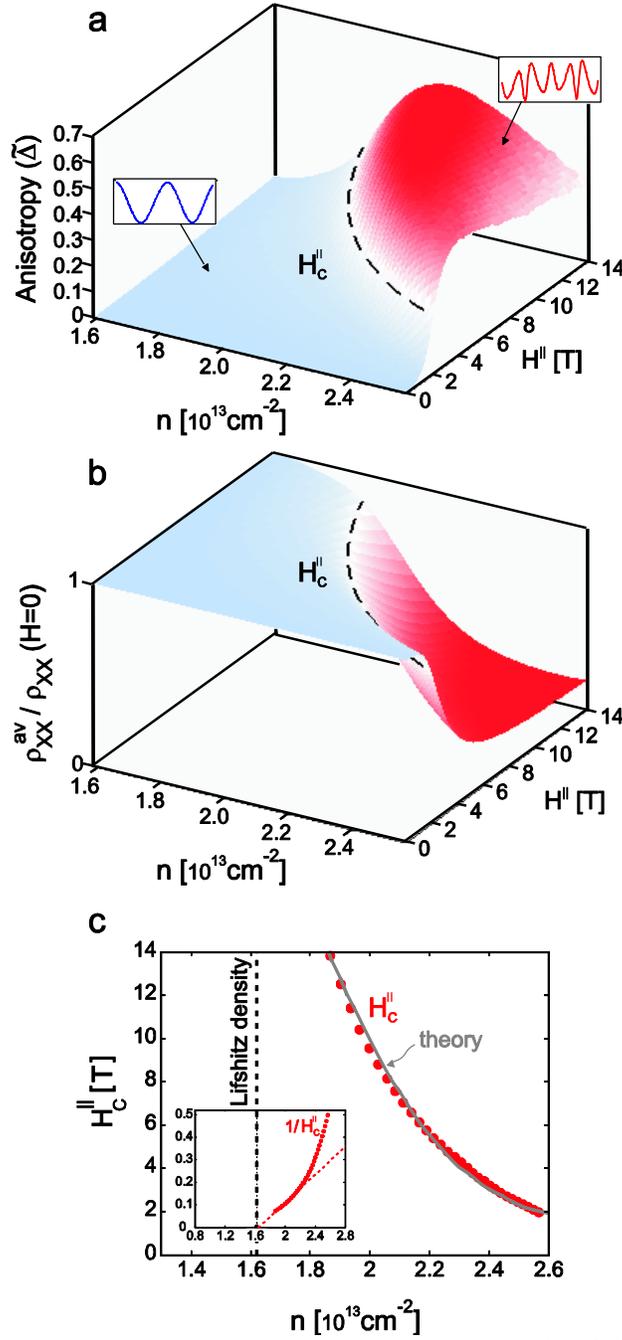

**Figure 3: Phase diagram in the density-field plane.** a) The magnitude of the anisotropy plotted as a function of the total carrier density and magnitude of the in-plane field. The anisotropy, $\tilde{\Delta} = (\rho_{XY}^{max} - \rho_{XY}^{min})/\rho_{XX}^{av}$, was determined by measuring for each density and field the modulation of $\rho_{XY}$ with $\phi_H$, extracting its peak-to-peak amplitude, $\rho_{XY}^{max} - \rho_{XY}^{min}$, and normalizing it by the average of $\rho_{XX}$ over the angle $\phi_H$, $\rho_{XX}^{av}$. A density-dependent critical field, $H_C^{\parallel}$ (dashed line) separates two regions of substantially different anisotropy magnitude and angular dependencies (indicated by the insets). b) The $\phi_H$-averaged $\rho_{XX}$ normalized to its value at $H^{\parallel} = 0T$, plotted in the same density-field plane. The indicated $H_C^{\parallel}$ (dashed line) is taken from panel a. c) The field, $H_C^{\parallel}$, extracted from panel a or b plotted vs. density (solid red circles).

The grey line is a fit to $H_C^{\parallel}$ based on the theoretical model (see Supplementary A2). The vertical dashed line is the Lifshitz critical density of the sample determined from perpendicular field measurements (22). The inset shows the density dependence of $1/H_C^{\parallel}$.

very similar to the scaling *perpendicular* field we reported elsewhere (22) (Supplementary Fig. S1). This empirical observation suggests that the effect of the magnetic field on transport, even in perpendicular fields, must involve spin-orbit interactions (Supplementary A1).

An important insight into the large-anisotropy phase is gained by tilting the field slightly out of plane ($\theta \approx 0.8°$). This is an unusual configuration to measure transport wherein along with the symmetric component we also measure an anti-symmetric (Hall) component of the transverse resistivity, $\rho_{XY}^A$ (Fig. 4a), which is strongly influenced by the dominant in-plane field. This anti-symmetric component is linear at low values of the total field, $H_{tot}$, around $H_C^{\parallel}$ it unexpectedly rises and then finally settles, at higher fields, on a slope comparable yet slightly smaller than that at low fields. As a function of the tilt angle, the low-field slope, $d\rho_{XY}^A/dH_{tot}$, scales perfectly as $\sin(\theta)$ all the way from in-plane to perpendicular field (inset of Fig. 4a). Thus, for $H < H_C^{\parallel}$, the linear dependence of $\rho_{XY}^A$ is simply due to the normal Hall effect induced by the perpendicular field component.

The surprising feature in the above measurement is the sharp rise of $\rho_{XY}^A$ near $H_C^{\parallel}$. If this was due to a normal Hall effect it would imply a rapid decrease in carrier density. However, judging from slopes of the linear regions below and above $H_C^{\parallel}$ it seems that the opposite happens, the density in fact slightly increases above $H_C^{\parallel}$. A more plausible origin of the sharp increase in $\rho_{XY}^A$ is an AHE due to the emergence of magnetization in the system. This unusual AHE is distinct from the "usual" AHE in the perpendicular configuration reported in the literature (27, 34). Compared to the usual AHE where the magnetization increase commences around zero field (35), here the effect appears suddenly around $H_C^{\parallel}$ behaving as a meta-magnetic transition. We note that this transition shows no evidence of a first-order discontinuity that characterizes conventional meta-magnetic transitions. Furthermore, this metamagnetic AHE is revealed only by suppressing the strong orbital effects present in the perpendicular configuration which cause a nonlinear HE unrelated to magnetization in the LAO/STO system (22, 28).

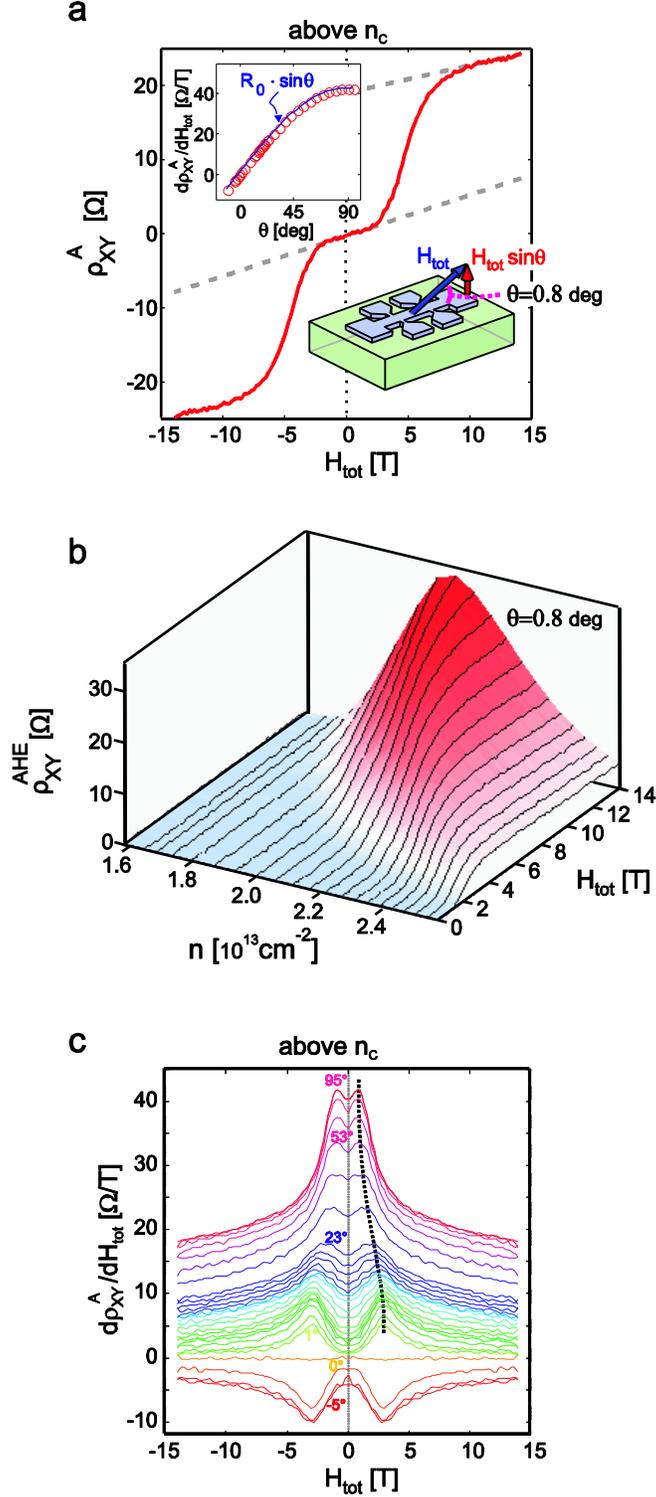

**Figure 4: Observation of an unusual Anomalous Hall Effect (AHE) above the Lifshitz transition.** a) The anti-symmetric component of the transverse resistivity, $\rho_{XY}^A(H_{tot}) = \left(\rho_{XY}(H_{tot}) - \rho_{XY}(-H_{tot})\right)/2$, vs. the magnetic field $H_{tot}$, applied at $\theta = 0.8°$ with respect to the plane of the interface. Dashed lines are guides to the eye to indicate the linear behavior at small and large fields. The low-field slope of the transverse resistance, $d\rho_{XY}^A/dH_{tot}$ (shown by the lower dashed line in panel a), as a function of $\theta$ (red circles in inset), is simply proportional to the component of the field out of the plane, $H_{tot}\sin\theta$, with a coefficient $R_0 = 42\ \Omega/T$ (fit shown by blue line in inset). b) The AHE component which develops at non-zero fields vs. density and field, obtained by subtracting out the low-field normal Hall slope, $\rho_{XY}^{AHE} = \rho_{XY}^{A} - H_{tot} \cdot d\rho_{XY}^{A}/dH_{tot}$. c) The derivative $d\rho_{XY}^{A}/dH_{tot}$ vs. $H_{tot}$ and $\theta$. The step in $\rho_{XY}^{A}$ due to the AHE at $H_C^{tot}$ shows up as a peak in $d\rho_{XY}^{A}/dH_{tot}$, prominent for nearly in-plane fields ($\theta \to 0°$) but surviving even when the field is applied out-of-plane ($\theta \to 90°$). The dashed line follows the evolution of this peak.

In Figure 4b we isolate the metamagnetic AHE component (see caption) and plot it over the entire field–density phase diagram. Interestingly, this AHE appears in perfect correlation with the large crystalline anisotropy (Fig. 3a) and the huge drop in resistivity (Fig. 3b). The appearance of the metamagnetic AHE suggests that an internal spin polarization develops for $H > H_C^{\parallel}$, which is converted to an anomalous hall component through spin-orbit coupling. The magnitude of this AHE increases together with $H_C^{\parallel}$ as the density is lowered toward $n_C$. This observation is consistent with increased spin-orbit coupling seen upon lowering the density (4), which we attributed to the orbital degeneracy at the Lifshitz transition (22).

Finally, we show that signatures of the metamagnetic AHE exist *even* for perpendicular fields. In perpendicular fields, the strong normal Hall signal masks this AHE, making it harder to detect. However, this AHE is clearly visible in the derivative $d\rho_{XY}^{A}/dH_{tot}$ (Fig. 4c), where the step in $\rho_{XY}^{A}$ shows up as a peak that is seen for the full range of angles $0° < \theta < 90°$. In our previous work, we consistently observed this peak at small perpendicular fields and noted that it could not be explained by two-band physics. The new data shown here identifies this peak with the metamagnetic AHE, which indeed goes beyond the simple band picture.

We now turn to discuss the nature of the two regimes observed in transport. It is tempting to associate the change in symmetry and magnitude of the AMR around $n_c$ solely to the onset of the occupation of the anisotropic $d_{xz}/d_{yz}$ bands. However, such a single-particle picture cannot account for the pinning of AMR along diagonal directions, the square-wave behavior of $\rho_{XY}$ and the existence of a critical field at which the AMR, AHE and $\rho_{XX}$ sharply change. A more plausible scenario involves also local magnetic moments whose easy-axes and scattering of itinerant electrons lead to crystalline AMR. However, in such a "magnetic semiconductor picture" the crystalline AMR appears at low fields and is suppressed for fields exceeding the scale of the anisotropic magnetic couplings responsible for their easy axes (32, 36, 37), whereas we see that crystalline AMR set in only *above* a critical field. Thus, this model does not explain why spin polarization appears only above a critical field and why the drop of resistivity is so large.

A possible explanation is that compared to magnetic semiconductors, here the local moments freeze into a glassy phase resulting in a critical field for their polarization. Random spin orientation which generates strong scattering in the magnetic channel is eliminated when the moments are polarized, possibly accounting for the observed large resistivity drop. On the other hand, within this picture we cannot easily understand the strong density-dependence of the critical field. In fact, magnetic domains observed in the LAO/STO system (9, 10) are density-independent (11), in contrast to the tunable polarization we find, and also vanish in patterned samples (9) such as are used in our experiments. In addition, a spin glass is expected to give rise to a hysteretic behavior in magnetic field, which we do not observe. Another appealing explanation may involve a spin-spiral phase (38), whose axis may be aligned with the magnetic field giving rise to AMR. This model too cannot, however, naturally explain the striking density dependence of the critical field.

Having excluded alternative scenarios, we show below that the best explanation for the counter-intuitive behavior of the data has to involve the $d_{XY}$ and $d_{xz}/d_{yz}$ itinerant electrons having competing couplings to the local moments. The moments themselves can be considered to have $d_{XY}$ character, as suggested by current theories of their origin based on charge ordering (14) or oxygen vacancy mechanisms (39). From symmetry arguments we show (Fig. 5) that these moments couple antiferromagnetically to the $d_{XY}$ electrons and ferromagnetically to the $d_{XZ}/d_{YZ}$ electrons (see also Supplementary A3). Such couplings lead to a competition between two phases: Below $n_C$, when only the $d_{XY}$ band is occupied, the moments are screened by their Kondo coupling to these electrons. Within this picture involving strong Kondo correlations between the itinerant electrons and the local moments (see also ref. (40)), exceeding the critical field is responsible for breaking the Kondo singlets (41) and for the polarization of the moments. Above $n_C$ the increasing occupation of the $d_{XZ}/d_{YZ}$ bands results in a competing ferromagnetic Hund's coupling that leads to a continuous drop of the critical field. Comparison of the critical field computed based on this model with the measured value reproduces well the density dependence of the critical in-plane field observed in the experiment and is shown in Fig. 3c (details in Supplementary A2). This picture provides a unified explanation for the concurrent changes observed in various transport properties across $H_C^\parallel$: Below $H_C^\parallel$, the moments are screened and thus act as unitary scatterers leading to high resistivity, no polarization, and simple anisotropy. Above $H_C^\parallel$, the moments get polarized and their scattering cross-section drops sharply leading to a low resistivity polarized state with crystalline anisotropy. The easy axes of this polarized state, reflecting the anisotropy in the g-factor for the coupling of the field to the moments, will eventually be overridden by intense enough fields yielding once again the original non-crystalline AMR. We note that a possible criticism of the Kondo picture is that it requires the concentration of impurities to be smaller or equal to the itinerant electron density, whereas a large concentration of paramagnetic moments was observed (9). However, the measured 1/T dependence of their susceptibility (9) suggests that the majority of moments are in fact free, and only a small fraction is coupled to the itinerant electrons. Indeed, recent experiments (42, 43) estimate them to have a significantly smaller density than that of the itinerant electrons. This lends further support to the Kondo model.

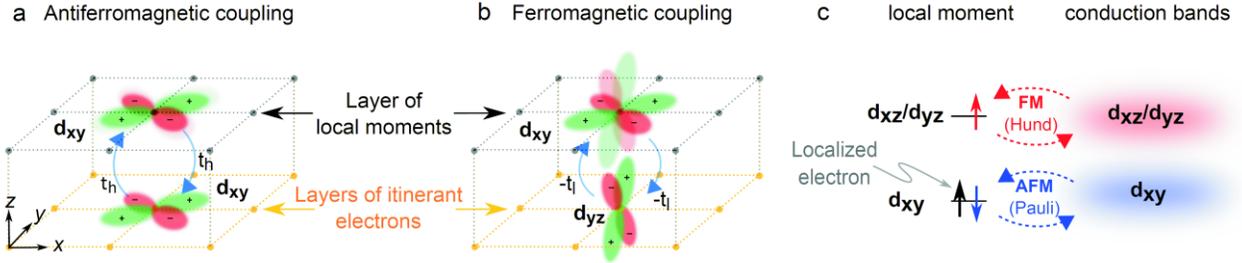

**Figure 5: Competing magnetic couplings between $t_{2g}$ conduction electrons and $Ti^{3+}$ local moments.** In our model $Ti^{3+}$ ions close to the interface form local magnetic moments of $d_{XY}$ symmetry. Subsequent $TiO_2$ layers, further away from the interface, harbor the itinerant electrons of either $d_{XY}$, $d_{XZ}$ or $d_{YZ}$ symmetry. a) The $d_{XY}$ itinerant spin hops into the occupied $d_{XY}$ state on the local moment and back. Here the conduction and local spins must be anti-aligned due to Pauli exclusion, resulting in an overall antiferromagnetic coupling. b) On the other hand, for the $d_{YZ}$ itinerant spin which hops into the unoccupied $d_{YZ}$ state on the moment site, a parallel alignment of spins is favored due to Hund's coupling on the local moment site. A similar ferromagnetic coupling is also favored for the $d_{XZ}$ itinerant electrons. c) A schematic diagram of the energy spectrum on the local moment site including the virtual processes giving rise to the competing magnetic couplings.

In Summary, AMR and AHE measurements in a planar field configuration show that the electronic system at the LAO/STO interface transitions at a critical magnetic field between two regimes with dramatically different anisotropy, polarization, and longitudinal resistivity. The clear density dependence of the critical field means that the itinerant electrons play an important role in the formation of these phases. This is surprising because the magnetic signatures of the LAO/STO system have so far been supposed to arise only from the local moments (whose origin is still debated). Our results not only provide compelling evidence for strong coupling between the itinerant electrons and moments, modeled to be localized in $d_{XY}$ orbitals at the interface (14, 15, 39), but also shed light on the symmetry-dependent nature of this coupling. This sets the stage for studying novel effects in the interacting system of moments and electrons at the LAO/STO interface where the polarization and easy axes develop only at high fields in contrast with conventional magnetic systems. The interplay between competing magnetic couplings studied here opens prospects for tunability by a gate of magnetism at the LAO/STO interface.

**Acknowledgements:** We would like to acknowledge A. D. Caviglia, S. Gariglio, A. Fete and J. –M. Triscone for the samples and fruitful discussions. We benefited greatly also from discussions with E. Berg, Y. Dagan, S. Finkelstein, D. Goldhaber-Gordon, Y. Meir, Y. Oreg, D. Shahar, A. Stern, V. Venkataraman and A. Yacoby. S.I. acknowledges the financial support by the ISF Legacy Heritage foundation, the Minerva foundation, the EU Marie Curie People grant (IRG) and the Alon fellowship. S.I. is incumbent of the William Z. and Eda Bess Novick career development chair. E.A. acknowledges the financial support by the ISF foundation. E.A. is incumbent of the Louis and Ida Rich Career Development Chair.

# SUPPLEMENTARY INFORMATION

## A Gate-tunable Polarized Phase of Two-Dimensional Electrons at the LaAlO$_3$/SrTiO$_3$ Interface

Arjun Joshua, J. Ruhman, S. Pecker, E. Altman and S. Ilani

*Department of Condensed Matter Physics, Weizmann Institute of Science, Rehovot 76100, Israel.*

A1. **The nature of the coupling to the magnetic field.**

A2. **Mean field theory of the t$_{2g}$ itinerant bands coupled to localized spins.**

A3. **Why $d_{XY}$ conduction electrons couple anti-ferromagnetically to localized moments whereas $d_{XZ}/d_{YZ}$ conduction electrons couple ferromagnetically to them.**

A4. **The detailed gate dependence of Hall resistivity for the sample in the main text.**

A5. **Phase diagram for a 10uc sample and another 6uc device.**

A6. **Methods**

**A1. The nature of the coupling to the magnetic field.**

In the main paper, we pointed out that the coupling of the applied magnetic field to the system must necessarily involve spin-orbit interactions. In this section, we expand more on the underlying reasoning.

In general the coupling of the field could be purely to spin, purely to orbital, or to a spin-orbit coupled system. Pure spin coupling can be ruled out immediately in our measurements since it would be independent of the direction of the in-plane field, $H^{\parallel}$, and so cannot cause AMR. Pure orbital coupling, on the other hand, seems to nicely fit some of the observations: This coupling has a natural critical field, the field at which the magnetic length is comparable to the confinement width of the 2D electrons. Besides, the $d_{XZ}/d_{YZ}$ wavefunctions being less confined than the $d_{XY}$ wavefunction (they are lighter in this direction) would explain why the coupling is better to them. Furthermore, orbital coupling would differentiate between the $d_{XZ}/d_{YZ}$ bands: Since orbital coupling is inversely proportional to the band mass perpendicular to the direction of $H^{\parallel}$, which for a general angle of the field in the plane, $\phi_H$, is not identical for the two bands, this coupling will lift their degeneracy and will induce orbital polarization making one Fermi ellipse larger than the other. This orbital polarization would lead to crystalline anisotropy, and if it also causes interband scattering to be suppressed, this scenario may also explain the drop in $\rho_{XX}$. However, despite this apparent agreement, three pieces of data exclude orbital coupling as the relevant mechanism. First, we observe that the $\rho_{XX}$ fall occurs for all angles $\phi_H$ at the same value of $H^{\parallel}$. If this drop was due to band polarization, no drop should have been observed along e.g. $\phi_H = 45°$, where $H^{\parallel}$ couples identically to both orbitals and does not lift their degeneracy. Second, orbital coupling cannot explain why the behavior is similar in parallel and perpendicular fields (Fig. S1, see also Ref. (1)). Lastly, the sign of orbital coupling is wrong: such coupling would increase the band mass perpendicular to the field direction, and thus $\rho_{XX}$ will increase when the field is perpendicular to the current, whereas we measure a decrease of $\rho_{xx}$ in this relative orientation (positive AMR). In fact, negative AMR was observed at elevated temperatures (2) suggesting that at high temperatures the width of the 2D is large and orbital effects are important. We observe the opposite sign at $T = 2K$, clearly showing that this is not an orbital effect. Thus, the only remaining coupling mechanism is that of a field to a spin-orbit coupled system.

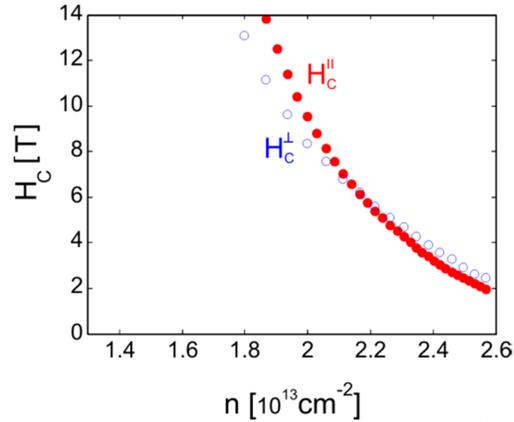

**Figure S1: Comparison of parallel and perpendicular critical fields.** $H_C^{\parallel}$ taken from Fig. 3c (solid red circles) and scaling perpendicular field (blue circles), reported previously (1), bear remarkable similarity both in their magnitude and in their trend as a function of total carrier density in the system.

**A2. Mean field theory of the $t_{2g}$ itinerant bands coupled to localized spins.**

Our starting point for the theoretical analysis is the Hamiltonian of three itinerant bands coupled to a lattice of localized moments

$$\mathcal{H} = \sum_{k} \sum_{\alpha,\alpha'=1}^{3} (\epsilon_{\alpha\alpha'}(\boldsymbol{k}) - \mu\, \delta_{\alpha\alpha'}) c^+_{\alpha k} c_{\alpha' k} + \mathcal{H}_{SO} + \mathcal{H}_H - \lambda \sum_i d^+_i d_i$$

$$+ J_K \sum_i \boldsymbol{S}_i \cdot \boldsymbol{s}_{1\,i} - J_H \sum_i \sum_{\alpha=2}^{3} \boldsymbol{S}_i \cdot \boldsymbol{s}_{\alpha\, i}$$

(Eq. S1)

The three bands $\alpha = 1,2,3$ represent the $d_{XY}$, $d_{XZ}$, and $d_{YZ}$ orbitals arranged in a square lattice slightly below the interface. The itinerant electrons interact with a lattice of localized electron spins, believed to reside on $d_{XY}$ orbitals of Ti near the interface layer (3, 4). The index $i$ represents the local moment sites, whose concentration we take as a phenomenological parameter. The spin operators $\boldsymbol{S}_i \equiv \frac{1}{2} d^+_i \boldsymbol{\sigma}\, d_i$ represent the local moment spins and $\boldsymbol{s}_i \equiv \frac{1}{2} c^+_{\alpha i} \boldsymbol{\sigma} c_{\alpha i}$ are the spin operators constructed from the itinerant electrons. Here $\boldsymbol{\sigma}$ are Pauli matrices acting in the electron spin space while $d_i$ and $c_{\alpha i}$ are two component spinor operators in this space representing the localized and itinerant electrons respectively. The chemical potential $\mu$ sets the density of the itinerant electrons and $\lambda$ is a Lagrange multiplier that fixes the density of the localized electrons.

The dispersion matrix $\epsilon(\boldsymbol{k})$ is the same as considered in the supplement to Ref. (1)

$$\epsilon(\boldsymbol{k}) = \begin{pmatrix} \frac{k^2}{2m_l} - \Delta_E & 0 & 0 \\ 0 & \frac{k_x^2}{2m_l} + \frac{k_y^2}{2m_h} & -\Delta_d k_x k_y \\ 0 & -\Delta_d k_x k_y & \frac{k_x^2}{2m_h} + \frac{k_y^2}{2m_l} \end{pmatrix}$$

where $m_l = 0.7\, m_e$ and $m_h = 15\, m_e$. In addition to nearest neighbor hopping we have included a diagonal hopping term $\Delta_d = m_h^{-1}$ that couples the $d_{XZ}$ and $d_{YZ}$ orbitals. $\Delta_E = 47$ meV is the energy offset of the $d_{XY}$ band. The atomic spin-orbit coupling is described by the local quadratic Hamiltonian

$$\mathcal{H}_{SO} = \Delta_{SO} \sum_{j,\alpha,\alpha'} c^+_{\alpha j} \boldsymbol{L}_{\alpha\alpha'} \cdot \boldsymbol{\sigma}\, c_{j\alpha'}$$

where $\boldsymbol{L}_{\alpha\alpha'}$ are the $l = 2$ angular momentum matrices projected to the space of the three $t_{2g}$ orbitals (see also the supplementary material in Ref. (1)). Finally, the coupling to an external in-plane magnetic field is given by the Hamiltonian

$$\mathcal{H}_H = -\mu_B\, \boldsymbol{H} \cdot \sum_j c^+_{\alpha j} (\boldsymbol{L}_{\alpha\alpha'} \otimes I + gI \otimes \boldsymbol{\sigma})\, c_{j\alpha'}$$

The first term above is the coupling of the magnetic field to the orbital angular momentum, the second is the Zeeman term, $I$ represents a unit matrix either in spin or orbital space and $g$ is the bare electronic g-factor.

To understand the quantum ground states of this model we employ a variational mean-field approximation in the spirit of the standard large $N$ mean-field theory of the Kondo lattice (see for example Ref. (5)). The variational wavefunction is generated from a quadratic mean-field Hamiltonian

$$\mathcal{H}_{MF} = \mathcal{H}_0 + \sum_{i,\sigma}(\chi\, d^+_{\sigma i} c_{\sigma 1 i} + h.c.) - \sum_j \sum_{\alpha=1}^{3} M_\alpha\, s_j^x - \sum_i M_d\, S_i^x$$

(Eq. S2)

Here $\mathcal{H}_0$ includes all the quadratic terms in Eq. S1. $\chi$ is the singlet hybridization field, which describes collective screening of the moment spins (6). The parameters $M_\alpha$ and $M_d$ account for the induced magnetization on the itinerant bands and the local moments respectively. We assume $M_2 = M_3$ to preserve orbital symmetry.

We solve for the variational parameters by minimizing the expectation value of the full Hamiltonian (Eq. S1) with respect to $\chi$, $M_1$, $M_2$ and $M_d$, where the expectation value is taken with the ground state of the mean-field Hamiltonian (Eq. S2). In other words, we seek the solution of the following set of equations

$$\frac{\partial \langle \mathcal{H} \rangle}{\partial \chi} = 0 \, ; \, \frac{\partial \langle \mathcal{H} \rangle}{\partial M_1} = 0 \, ; \, \frac{\partial \langle \mathcal{H} \rangle}{\partial M_2} = 0 \, ; \, \frac{\partial \langle \mathcal{H} \rangle}{\partial M_d} = 0 \qquad \text{(Eq. S3)}$$

These equations are supplemented by two additional equations

$$-\frac{\partial \langle \mathcal{H} \rangle}{\partial \mu} = \sum_j \sum_{\alpha=1}^{3} \langle c^+_{\alpha j} c_{\alpha j} \rangle = N \quad ; \quad -\frac{\partial \langle \mathcal{H} \rangle}{\partial \lambda} = \sum_i \langle d^+_i d_i \rangle = N_d$$

which fix the number of itinerant and localized electrons independently.

In the main paper, we argued that the diverging critical field obtained from this model (grey line in Fig. 3c) is consistent with a transition from a Kondo phase to a magnetically polarized phase. Indeed, the set of variational equations (Eq. S3) have two distinct solutions: One where $\chi$ is finite which is identified with the Kondo or "heavy liquid" phase, and the other where $\chi$ is strictly zero and the moments are fully polarized. The theoretical fit to the critical field presented in Fig. 3c is obtained by comparing the energy of these two variational wavefunctions as a function of the applied field $H$, using $n_d = 2.6 \times 10^{12}$ cm$^{-2}$, $J_K = 900$ meV, and $J_H = 625$ meV. This approach predicts that the transition is of first order, that is, the value of $\chi$ at the transition is finite (see Fig. S2). Generically this should be expected because both phases do not break any symmetry of the Hamiltonian (Eq. S1). However it should also be pointed out that large $N$ mean-field theories are known to give spurious first order transitions.

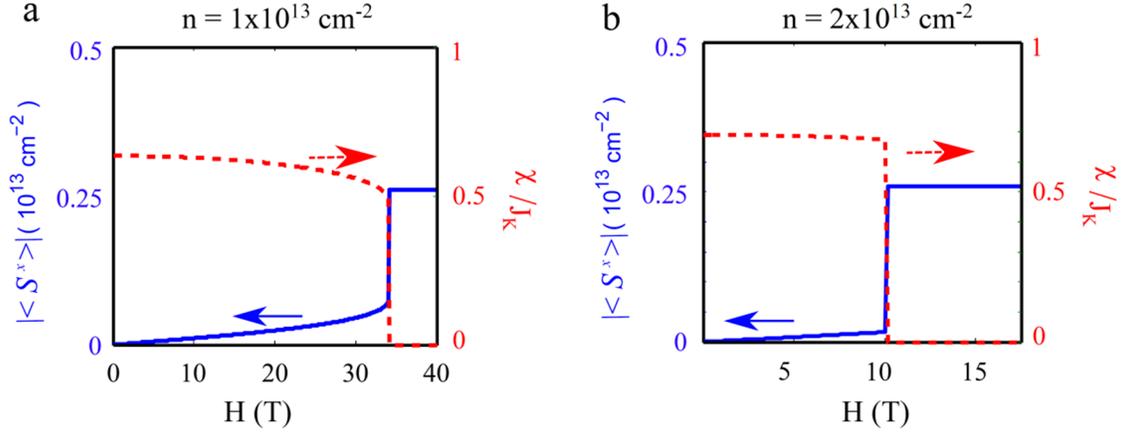

**Figure S2: Magnetic field driven transition from "heavy liquid" phase to polarized phase, given by the variational calculation.** The average magnetization of the moment band $|\langle S^x \rangle|$ and the singlet hybridization $\chi/J_K$ plotted as a function of the applied magnetic field $H$ for a) itinerant electron density $n = 1 \times 10^{13}\,\text{cm}^{-2}$ (below the Lifshitz density) and b) $n = 2 \times 10^{13}\,\text{cm}^{-2}$ (above the Lifshitz density). At a critical value $= H_c$, $\chi$ jumps to zero whereas $\langle S^x \rangle$ jumps to its maximum value. For both panels, the parameters are $n_d = 2.6 \times 10^{12}\,\text{cm}^{-2}$, $J_K = 900$ meV, and $J_H = 625$ meV.

## A3. Why $d_{XY}$ conduction electrons couple anti-ferromagnetically to localized moments whereas $d_{XZ}/d_{YZ}$ conduction electrons couple ferromagnetically to them.

To understand why the $d_{XY}$ conduction electrons couple differently to localized moments compared to the $d_{XZ}/d_{YZ}$, let us consider the on-site Hamiltonian of a single local moment:

$$\mathcal{H}_d = \sum_{\alpha} \epsilon_\alpha n_\alpha^d + U \sum_\alpha n_{\alpha\uparrow}^d n_{\alpha\downarrow}^d + U' \sum_{\alpha \neq \alpha'} \sum_{\sigma\sigma'} n_{\alpha\sigma}^d n_{\alpha'\sigma'}^d - J \sum_{\alpha\alpha'} \mathbf{S}_\alpha \cdot \mathbf{S}_{\alpha'}.$$

Here $n_\alpha^d$ is the density operator of the localized state, $\epsilon_\alpha$ are the single particle energies, $U$ and $U'$ are the inter- and intra-orbital Hubbard interactions and $J$ is the Hund rule's coupling.

The energies $\epsilon_\alpha$ belong to the different orbital states on the moment: $\epsilon_1$ belongs to the $d_{XY}$ orbital and $\epsilon_2, \epsilon_3$ are the energies of $d_{XZ}, d_{YZ}$ (Fig. S3d). The splitting between these states $\delta = \epsilon_2 - \epsilon_1$ is unknown and might be large (3).

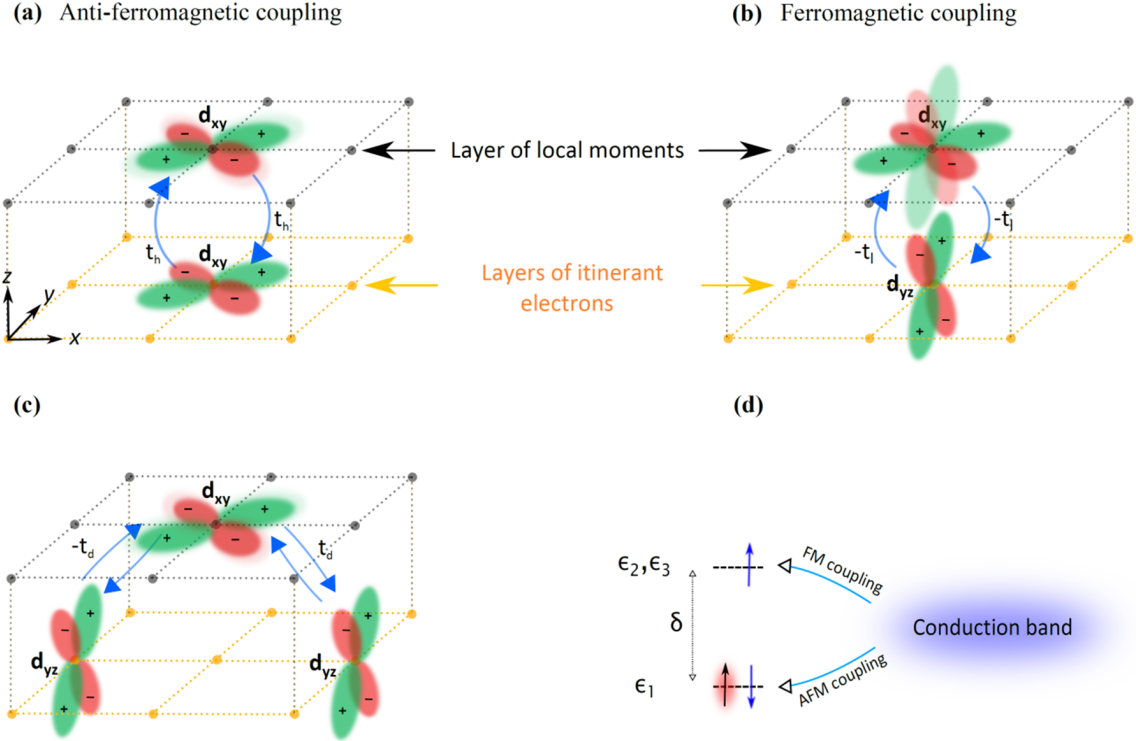

**Figure S3: Schematic representation of the hopping processes between localized and itinerant electronic states.** The first few TiO$_2$ layers are occupied by the localized states of $d_{XY}$ symmetry whereas subsequent TiO$_2$ layers, further away from the interface, hold itinerant electrons that can be of either $d_{XY}$, $d_{XZ}$ or $d_{YZ}$ symmetry. a) Hopping between the $d_{XY}$ itinerant electron and a $d_{XY}$ localized state. This process has a small "heavy" hopping amplitude, $t_h$, since the lobes of the wavefunctions are perpendicular to the hopping direction. b) Hopping between the $d_{YZ}$ itinerant electron and the unoccupied $d_{YZ}$ state on the moment site. This process has a large "light" amplitude $t_l$, since the lobes of the wavefunctions are pointing in the hopping direction. c) Hopping between the $d_{YZ}$ itinerant electrons and the $d_{XY}$ localized state is only allowed through a next-nearest-neighbour diagonal hopping process (1), with amplitude $t_d$ that is comparable to the heavy hopping amplitude. d) The resulting exchange coupling. The energy diagram of the moment state is shown, whose occupied $d_{XY}$ state is lower in energy by $\delta$ compared to the unoccupied $d_{XZ}$ and $d_{YZ}$ states. The process in a) gives an AFM superexchange between itinerant $d_{XY}$ electrons and the localized moment. The virtual hopping process in b) together with local Hund's coupling on the localized state yield an effective FM coupling between $d_{XZ}$ and $d_{YZ}$ electrons and the localized moment. The latter process has a large energy denominator ($\delta$), but on the other hand involves much larger hopping amplitude $t_l \gg t_h, t_d$. Since the exchange terms are quadratic in the hopping amplitudes this term would be quite large.

The effective magnetic coupling between conduction electrons and the localized moments can be estimated in second order perturbation theory where the itinerant electron hops into a virtual state on the localized site and back (see Fig. S3). The hopping along the $z$-direction of a $d_{XY}$ itinerant electron into a $d_{XY}$ localized state on the moment has a small ("heavy") hopping amplitude, $t_h$ (Fig. S3a), resulting in the following hybridization Hamiltonian

$$\mathcal{H}_{hyb}^1 = -t_h \sum_i (d_{1i}^+ c_{1i} + h.c.).$$

This is exactly the hopping element responsible for the heavy electronic mass in STO, and its smallness results from the small overlap of the $d_{XY}$ wavefunctions along the $z$-direction (their lobes are in the $XY$ plane). This process is possible only if the hopping is to a state with spin anti-parallel to the spin of the localized moment. The intermediate energy of this second order process is approximately $U + \epsilon_1$. Similarly, the localized electron can hop into the conduction band and back with an intermediate energy denominator $\sim \epsilon_1$, therefore the coupling is anti-ferromagnetic and is given by (6):

$$J_K \sim t_h^2 \left( \frac{1}{U + \epsilon_1} - \frac{1}{\epsilon_1} \right) \qquad \text{(Eq. S4)}$$

For the $d_{XZ}/d_{YZ}$ conduction electrons one can think of two processes, the first is where the conduction electrons hop into higher unoccupied states at the moment's site with the same orbital symmetry (Fig. S3b), which is described by the following hybridization Hamiltonian

$$\mathcal{H}_{hyb}^{23} = -t_l \sum_i \sum_{\alpha=2}^3 (d_{\alpha i}^+ c_{\alpha i} + h.c.).$$

In this case electrons with both parallel and antiparallel spin with respect to the spin of the localized moment can hop in. However, due to the Hund's coupling, the hopping to states with parallel spin alignment will have a smaller energy denominator $U' + \epsilon_2 - J$, as compared to hopping to states with anti-parallel spin alignment, whose energy denominator is $U' + \epsilon_2 + J$. The resulting effective coupling between the $d_{XZ}$, $d_{YZ}$ itinerant electrons and the localized moment is therefore ferromagnetic and is given by:

$$J_H \sim t_l^2 \left( \frac{1}{U' + \epsilon_2 - J} - \frac{1}{U' + \epsilon_2 + J} \right) \qquad \text{(Eq. S5)}$$

The $d_{XZ}/d_{YZ}$ itinerant electrons can also hop to a $d_{XY}$ state on the localized site through a next-nearest-neighbor diagonal hopping processes with amplitude $t_d$ (Fig. S3c)

$$\widetilde{\mathcal{H}}_{hyb}^{23} = -t_d \sum_{\langle ii' \rangle} \sum_{\alpha=2}^{3} (\eta_{ii'}^\alpha \, d_{\alpha i}^+ c_{\alpha i'} + h.c.) \, .$$

where $\eta_{ii'}^\alpha = \text{sign}(i - i')$ and is non-zero only if $i - i'$ is along $y$ and $\alpha = 2$ or along $x$ and $\alpha = 3$. This hybridization is the same as the diagonal hopping which hybridizes the $d_{XZ}$ and $d_{YZ}$ in the plane (see the supplement to Ref. (1)). Just as the process described in Fig. S3a, this process too will give rise to an AFM superexchange coupling. However, this coupling is negligible compared to the ferromagnetic coupling for two reasons. The first is that the hopping processes from the two sides of the moment have opposite signs and therefore interfere destructively. The second is that the hopping element $t_d$ is an order of magnitude smaller than the light hopping $t_l$ responsible for the FM coupling in Fig. S3b.

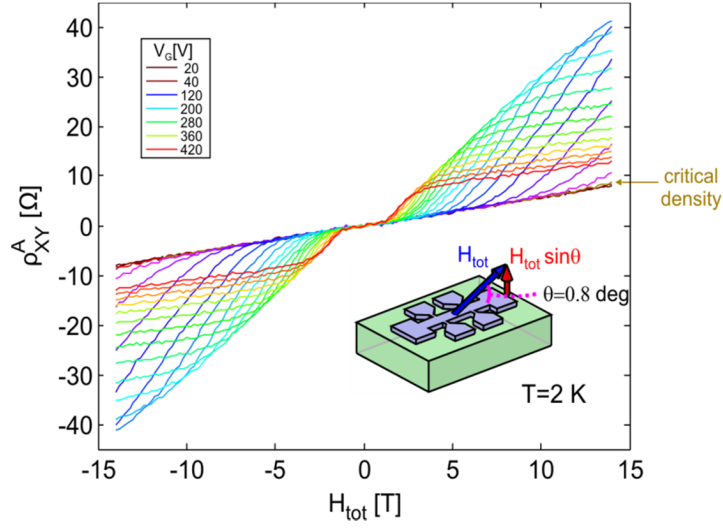

**Figure S4: Hall resistance, $\rho_{XY}^A$, as a function of nearly in-plane field, $H_{tot}$, for various gate voltages**. At densities much above the Lifshitz value (high gate voltages), the traces have a step at a critical field. The traces become completely linear as the density is tuned from above to below the Lifshitz critical density (low gate voltages). The inset shows the geometry of the measurement.

### A4. The detailed gate dependence of Hall resistivity for the sample in the main text.

In Fig. 4b of the main paper we showed, as a three-dimensional plot, the dependence of the anomalous Hall effect (AHE), $\rho_{XY}^{AHE}$, in the density-field plane after subtracting out the low-field slope from the measured anti-symmetric component of the transverse resistivity, $\rho_{XY}^A$. The AHE thus obtained, $\rho_{XY}^{AHE} = \rho_{XY}^A - H_{tot} \cdot d\rho_{XY}^A / dH_{tot}$, was negligibly small for low fields but showed a step that developed rapidly above a critical field. Here we show the raw dataset of $\rho_{XY}^A$ as a function of the total field, $H_{tot}$, and density without the subtraction procedure.

The various gate voltages used to tune the density correspond to the individual line traces in Fig. S4. At the highest gate voltages used so that the carrier density was well above the Lifshitz critical density, the $\rho_{XY}^A$ trace (red) is linear in $H_{tot}$ at low fields. Around $H_C^{\parallel} = 2.5T$ it rises sharply, and at higher fields it once again stabilizes on a gradual linear dependence on $H_{tot}$. The slope at these higher fields is comparable, yet slightly smaller than that at low fields. As the gate voltage is decreased and the density is lowered in the system, two changes are immediately apparent in the shape of the line trace: whereas the low-field slope remains almost fixed, the step in $\rho_{XY}^{A}$ occurs at a larger $H_{C}^{\parallel}$ and the size of the step is also larger. At fields $H_{tot} > H_{C}^{\parallel}$, once again the Hall resistivity $\rho_{XY}^{A}$ settles on a slope that is slightly smaller compared to its low-field value. These trends progressively continue as the gate voltage is decreased further (red through orange, green and cyan traces), until $H_{C}^{\parallel} > 14 T$ and the step lies outside our maximum applied field leaving only the initial rise of the step to be seen (blue to purple traces). Finally, the density falls just below the critical density (brown and black traces) and $\rho_{XY}^{A}$ remains completely linear up to the highest field. Thus, the build-up of the AHE at $H_{C}^{\parallel}$ and the divergence of this critical field at the Lifshitz density, characteristic features of the LAO/STO system reported in the main paper, are also clearly visible in the raw data.

A notable feature common to all the traces is that they ride on a slope that has approximately the same value for all of them. This observation can be traced to the fact that the slope of the Hall resistivity i.e. the Hall coefficient at low values of the perpendicular field is inversely proportional to the number of high-mobility $d_{XY}$ carriers (1). Due to the fact that the field is applied almost in-plane, its perpendicular component is quite small even up to the maximum applied value of $H_{tot} = 14T$. The overall slope of the traces shown in Fig. S4 is therefore determined by the density of carriers in the $d_{XY}$ band, which remains fixed at the critical density for all densities of the total number of carriers exceeding this value (red to purple traces). In fact, since the gate is not tuned significantly below the critical value, even the lowest density (brown and black) traces shown in Fig. S4 do not have a significantly larger slope compared to the high-density traces.

## A5. Phase diagram for a 10uc sample and another 6uc device.

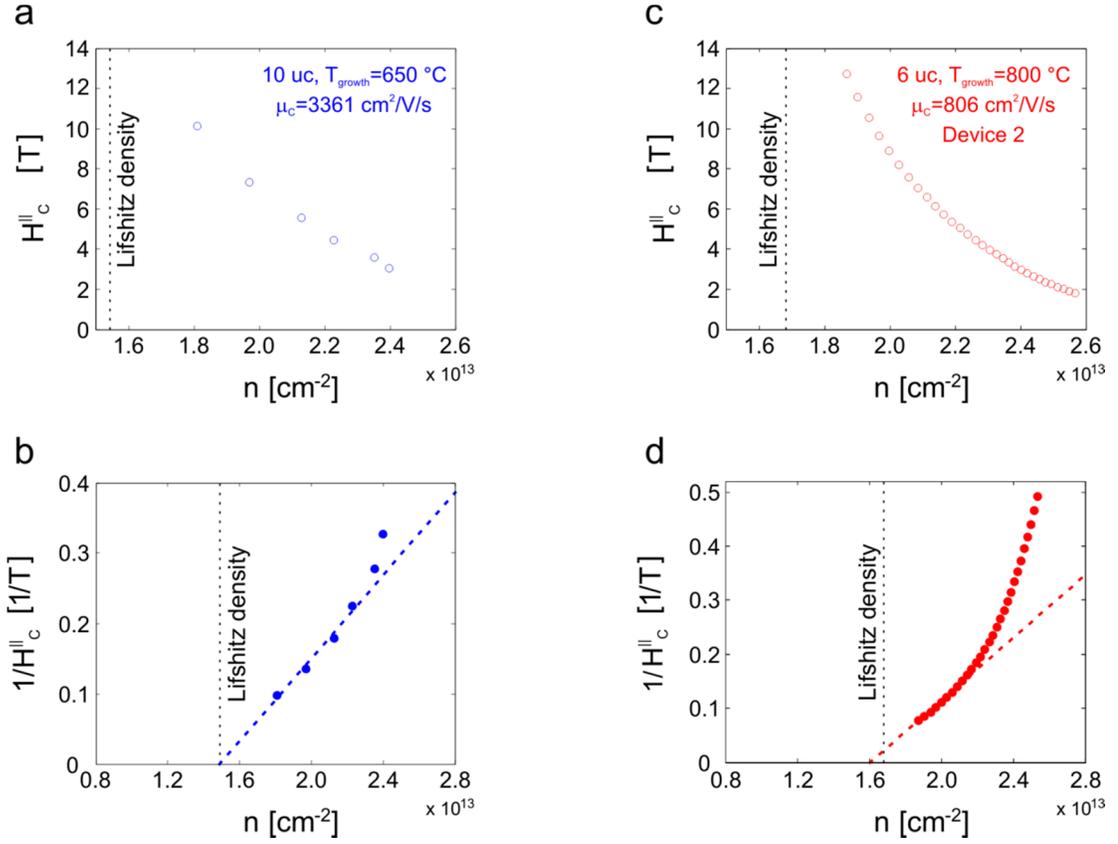

**Figure S5: Phase diagram for additional 10uc sample and another 6 uc device.** a) $H_C^{||}$ and b) $1/H_C^{||}$ extracted from anomalous Hall effect measurements of a 10uc sample, as a function of total carrier density. c) $H_C^{||}$ and d) $1/H_C^{||}$ from anisotropic magnetoresistance of another device on the 6 uc sample of the main paper, vs. density.

In the main paper, we presented anisotropic magnetoresistance (AMR) and anomalous Hall effect (AHE) data from a 6uc sample grown at $T_{growth}$=800 °C. Here we show similar results obtained in a 10uc sample, grown at a different temperature, $T_{growth}$=650 °C. Specifically, we show that the density dependence of the critical field extracted from AMR and AHE measurements across samples is similar.

Figure S5a shows the critical field, $H_C^{||}$, extracted from the position of the step in AHE measurements on a high mobility 10 uc sample. In these AHE measurements, the field was applied at an angle of $\theta$=0.7° to the plane of the interface. We find that $H_C^{||}$ increases as the LAO/STO system is progressively depleted. Plotting the reciprocal of this field, $1/H_C^{||}$, we find that at low densities, the data points collapse onto a single line that extrapolates to zero in the vicinity of the critical density associated with the Lifshitz transition (Fig. S5b). Exactly the same trends are also seen from AMR measurements on another Hall device on the same sample reported in the main paper: $H_C^{\parallel}$ appears to diverge at the critical density (Figs. S5c,d). Thus, as shown in the main manuscript, not only is the critical field determined from AMR and AHE similar (Figs. 3a and 4b), but this observation holds even across samples, where $H_C^{\parallel}$ appears to be divergent in the vicinity of the Lifshitz transition characteristic for each sample.

**A6. Methods.**

*Sample fabrication*

As detailed in earlier work (7), films were grown on $TiO_2$-terminated (001) $SrTiO_3$ single crystals of dimensions $5\,mm \times 5\,mm$ by pulsed laser deposition in $\sim 10^{-4}$ mbar of $O_2$. The repetition rate of the laser was 1Hz, with the fluence of each pulse being 0.6 J cm$^{-2}$. The film growth was monitored in situ using reflection high-energy electron diffraction. The 6uc (/10uc) sample was grown at T= 800 °C (T= 650 °C). After growth, the samples were annealed in 200 mbar of $O_2$ at about 600 °C (/530 °C) for one hour and cooled to room temperature in the same oxygen pressure. Hall bars were photolithographically patterned and the sample was ultrasonically bonded using Al wire.

*Measurements*

We used back-gated Hall bars with widths ranging from 100μm to 500um, oriented along the (100) crystallographic direction. Current (amplitude of 46 nA at frequencies ranging from DC to 13 Hz) was passed along this direction, and the longitudinal and transverse resistivities ($\rho_{XX}$ and $\rho_{XY}$) were measured while rotating the sample in a magnetic field applied in the plane of the interface at temperatures of T=2K. The absence of nonlinear effects was confirmed by ensuring similar data was measured despite lowering the amplitude of the current by an order of magnitude. The possibility of non-equilibrium effects was ruled out by testing different durations of wait-time after perturbing the system and subsequently using, in the data acquisition, a wait-time for which the system had relaxed.